\documentstyle[aps]{revtex}
\draft
\begin{document}
\title{Enhacement in the dymanic response of a viscoelastic fluid
flowing through a longitudinally vibrating tube}
\author{David Tsiklauri and Igor Beresnev}
\address{Department of Geological and Atmospheric Sciences,
Iowa State University, 253 Science I, Ames, IA 50011-3212,
U.S.A.\\
email: dtsiklau@iastate.edu  beresnev@iastate.edu}
\maketitle
\begin{abstract}
We analyzed effects of elasticity on the dynamics of fluids in porous
media by studying a flow of a Maxwell fluid in a tube,
which oscillates longitudinally and is subject to oscillatory
pressure gradient.
The present study investigates novelties
brought about into
the classic Biot's theory of
propagation of elastic waves in a fluid-saturated
porous solid
by inclusion of non-Newtonian effects
that are important, for example, for hydrocarbons.
Using the time Fourier transform and transforming the
problem into the frequency domain, we calculated: (A)
the dynamic permeability and
(B) the function $F(\kappa)$ that measures the deviation from
Poiseuille flow friction as a function of frequency
parameter $\kappa$.
This provides a more complete theory of flow of Maxwell
fluid through the longitudinally oscillating cylindrical
tube with the oscillating pressure gradient, which has important
practical applications.
This study has clearly shown transition from dissipative to
elastic regime in which sharp enhancements (resonances)
of the flow are found.
\end{abstract}
\date{\today}


\section{Introduction}

A quantitative theory of propagation
of elastic waves in a fluid-saturated
porous solid was formulated in the classic papers by Biot [1].
One of the major findings of Biot's work was
that there was
a breakdown in Poisseuille flow above a certain characteristic frequency

specific to the fluid-saturated porous material.
Biot theoretically studied this phenomenon by
considering the flow of a viscous fluid in a tube with
longitudinally oscillating walls under an oscillatory pressure
gradient. Apart from the fundamental interest, the investigation of the
dynamics of fluid in porous media under oscillatory pressure
gradient and oscillating pore walls
is of prime importance for the recently emerged technology
of acoustic stimulation of oil reservoirs [2].
For example,
it is known that the natural pressure in an oil reservoir
generally yields no more than approximately 10 percent oil recovery.
The residual oil is difficult to produce due to its naturally
low mobility, and the enhanced oil recovery operations are
used to increase production.  It has been experimentally proven
that there is a substantial increase in the net fluid flow
through porous space if the latter is treated with elastic waves.
However, there is a fundamental lack of
understanding of the physical mechanisms of fluid mobilization in
saturated rock through the effect of elastic waves;
the theory of such mobilization virtually does not exist.
Biot's theory can be used to describe the interaction of fluid-saturated

solid with the sound for a classic Newtonian fluid;
however, oil and other hydrocarbons exhibit significant
non-Newtonian behavior [3].
The aim of this paper is therefore to incorporate non-Newtonian effects
into the classical study of Biot [1].

Recently, del Rio, Lopez de Haro, Whitaker [4]
presented a
study of enhancement in the dynamic response of a viscoelastic
(Maxwell) fluid
flowing in a stationary (non-oscillating) tube under the effect of
an oscillatory pressure gradient.
We combine this theory with the effect of the acoustic
oscillations of the walls of a tube introduced by Biot [1],
providing a complete description of the interaction
of Maxwell fluid, filling the pores, with acoustic waves.

Finally, in order to emphasize that the
consept of dynamic permeability is an adequate way
of desciprtion of the phenomenon, we note that this
concept has been widely used before [5].

In the next section we formulate our model, whereas
the last section will conclude with the discussion of the results.

\section{The model}

In this section we present our model of Maxwell fluid flowing in
a cylindrical tube whose walls are oscillating longitudinally
and the fluid is subject to an oscillatory pressure gradient.
We give analitical solutions of the problem in the frequency domain.

The governing equations of the problem consist of the
continuity equation for the incompressible fluid,
$$
\nabla \cdot \vec v =0, \eqno(1)
$$
and the linearized momentum equation,
$$
\rho {{\partial \vec v}\over{\partial t}} = - \nabla p - \nabla \tilde
\tau.
\eqno(2)
$$
Here, $\vec v$, $p$, $\rho$ denote velocity, pressure and mass density
of the fluid, whereas $\tilde \tau$ represents the viscous stress
tensor.
We describe the viscoelatic effects of the fluid using Maxwell's
model, which assumes that
$$
t_m {{\partial \tilde \tau}\over{\partial t}}= -\eta \nabla \vec  v  -
\tilde \tau, \eqno(3)
$$
where $\eta$ is the viscosity coefficient and $t_m$ is the
relaxation time.

Now, let $u$ be a velocity of the wall of the tube which oscillates in
time
as $e^{- i \omega t}$, where $\omega$ is the angular frequency.
The flow of fluid in a cylindrical tube with longitudinally
oscillating walls can be descibed by a single component of
the velocity, namely, its $z$-component $v_z$ ($z$ axis is along
the ceterline of the tube). We use the cylindrical coordinate
system $(r,\phi,z)$ in the treatment of the problem.
We introduce the relative velocty $U_1$ as
$U_1=v_z-u$. Thus, assuming that all physical quantities vary in time as

$e^{- i \omega t}$, we arrive at the following master equation for $U_1$

$$
\nabla^2 U_1 +
{{\omega^2 t_m + i \omega}\over{\nu}} U_1=
 -{{X}\over{\nu}}(1 - i \omega t_m).
\eqno(4)
$$
Here, we have introduced the following notations:
$$
X= -(\nabla p + \rho{{\partial u}\over{\partial t}} ),
$$
which is a sum of the applied pressure gradient and force exerted
on the fluid from the oscillating wall of the tube and
$\nu$, which is $\nu= \eta / \rho$. Note that a no-slip boundary
condition at the wall is assumed.

The solution of Eq.(4) can be found to be [1]
$$
U_1 = - {{X}\over{i \omega}} + C J_0(\beta r),
$$
where $J_0$ is the Bessel function and
$\beta = \sqrt{(\omega^2 t_m + i \omega)/{\nu}}$.

Applying a no-slip boundary condition $U_1(a)=0$ at the
wall of the tube, where $a$ is its radius, we finally get
$$
U_1(r)= - {{X}\over{i \omega}}
\left[1 - {{J_0(\beta r)}\over{J_0(\beta a)}} \right]=
-{{Xa^2(1 - i \omega t_m)}\over{ \nu }} {{1}\over{(\beta a)^2}}
\left[1 - {{J_0(\beta r)}\over{J_0(\beta a)}} \right].
\eqno(5)
$$

Defining the cross-section averaged velocity as
$$
\bar U_1 ={{2}\over{a^2}} \int_0^a U_1(r) r d r,
$$
we obtain
$$
\bar U_1=-{{Xa^2(1 - i \omega t_m)}\over{ \nu }} {{1}\over{(\beta a)^2}}
\left[1 - {{2 J_1(\beta a)}\over{(\beta a) J_0(\beta a)}} \right]
\equiv K(\omega) X.
\eqno(6)
$$
Here $K(\omega)$ is the dynamic permeability [4] that describes the
frequency
dependent response of the tube to the applied total force on the fluid.
Simple comparison reveals that Eq.(6) resembles closely
Eq.(6) from Ref. [4], with the only difference being that
we have $X$ in place of their $\partial p / \partial z$.
Note that in the case of stationary tube walls
($\rho (\partial u/\partial t) \to 0$)  Eq.(6) exactly
coincides with  Eq.(6) from Ref.[4].
Simple calculation (applying L'Hospital rule for the $0/0$
uncertanty) shows that
$\lim_{\omega \to 0} K(\omega) =a^2 / (8 \nu)$.
Thus, following Ref.[4], we will introduce the
dimensionless dynamic permeability
as $K^*(\omega) = K(\omega) / K(0)$, which will be used later
(see Fig.7). Note, that we were able
to easily reproduce Fig. 1 from Ref. [4], confirming the existence
of sharp resonances of $K^*(\omega)$ in the elastic regime (see below
for the definition of this regime).

Following the work of Biot [1] we calculate the stress at the
wall $\tau$,
$$
\tau = -{{\eta}\over{1 - i \omega t_m}}
\left({{\partial U_1(r)}\over{\partial r}}\right)_{r=a}
= {{\eta \beta X}\over{i \omega (1 -i \omega t_m)}}
{{J_1(\beta a)}\over{J_0(\beta a)}}.
\eqno(7)
$$
Note, that when $t_m \to 0$ this expression
obviously coincides with the corresponding
Newtonian form.

The total friction force is $2 \pi a \tau$. Following
Biot we calculate the ratio of total friction
force to the average velocity, i.e.
$$
{{2 \pi a \tau}\over{\bar U_1}}=
- {{2 \pi \eta (\beta a)
[J_1(\beta a) / J_0(\beta a)]}
\over{(1 -i \omega t_m)}}
\left[1 - {{2 J_1(\beta a)}\over
{(\beta a) J_0(\beta a)}} \right]^{-1}.
\eqno(8)
$$
Simple analysis reveals that
$$
\lim_{\omega \to 0}
{{2 \pi a \tau}\over{\bar U_1}}=8 \pi \eta,
$$
which corresponds to the limiting case of
Poiseuille flow. Following Biot [1], we also introduce a function
$F(\kappa)$ with $\kappa$ being
$\kappa= a \sqrt{\omega / \nu}$ in the following
manner
$$
{{2 \pi a \tau}\over{\bar U_1}}=
8 \pi \eta F(\kappa),
$$
thus,
$$
F(\kappa)=-{{1}\over{4}}
{ { \kappa \sqrt{ i + \kappa^2 / \alpha}
\left[J_1(\kappa \sqrt{ i + \kappa^2 / \alpha})
/ J_0(\kappa \sqrt{ i + \kappa^2 / \alpha})\right]}
\over{(1 -i \kappa^2 / \alpha)}}
\left[1 - {{2 J_1(\kappa \sqrt{ i + \kappa^2 / \alpha})}\over
{\kappa \sqrt{ i + \kappa^2 / \alpha}
J_0(\kappa \sqrt{ i + \kappa^2 / \alpha})}}
\right]^{-1}.
\eqno(9)
$$
Note, that $F(\kappa)$  measures the deviation from
Poiseuille flow friction as a function of frequency parameter $\kappa$,
as introduced by Biot [1].

In Eq.(9), $\alpha$ denotes the Deborah number [4], which is defined as
the ratio of characteristic time of viscous effects $t_v=a^2/ \nu$ to
the
relaxation time $t_m$, i.e.
$\alpha= t_v / t_m =a^2/(\nu t_m)$.

As noted in Ref. [4], the value of the parameter $\alpha$
determines in which regime the system resides. Beyond
a certain critical value ($\alpha_c=11.64$), the system is
dissipative, and viscous effects dominate. On the other hand, for
small $\alpha$ ($\alpha < \alpha_c$) the system
exhibits viscoelastic behavoir which we call the
elastic regime.

Note, that the Newtonian flow regime can be easily recovered from
Eq.(9) by putting $\alpha \to \infty$.
We plot this limiting case in Fig. 1. As it can be
seen from the figure,  both $Re[F (\kappa)]$
and $Im[F (\kappa)]$ exactly coincide with the
Newtonian limiting case studied in Biot's work (see Fig. 4 in
Ref.[1]). This graph demonstrates a breakdown in Poiseuille
flow as frequency increases (recall that $\kappa
\propto \sqrt{\omega}$).
In all our calculations we have used polynomial
expansions of $J_0$ and $J_1$ with absolute error not
exceeding $10^{-6}$ \%. Thus, our calculation results are
accurate to this order.

A finite-but-large $\alpha$ regime is shown in the next two
figures. Fig. 2 corresponds to the case when $\alpha = 10^4$.
We see in Fig. 2 that the real and
imaginary parts of $F(\kappa)$ start to
deviate from the Newtonian fluid behavior at {\it large}
frequencies. In Fig. 3 solutions
correspond to the case when $\alpha = 100.0$; thus we see how
viscoelastic effects become pronounced already at {\it low} frequencies.

Fig. 4 presents the behavior of $Re[F (\kappa)]$
and $Im[F (\kappa)]$ when  $\alpha = 10.0$. As it can be seen from
the graph, sharp resonances appear on the curves. This feature can
be explained by the fact that in this case  $\alpha$ is
less than $\alpha_c$, which means that the system switched
from the dissipative (viscous) regime to the elastic one.

Fig. 5 shows the solutions when  $\alpha = 1.0$. In this case
we see more irregular behavior of $Re[F (\kappa)]$
and $Im[F (\kappa)]$ with a number of irregular spikes.

The extreme non-Newtonian (elastic) regime is studied in Fig. 6
where we plot the solutions for the case when
$\alpha = 10^{-3}$. In this case notable change is that
there are fewer but more pronounced spikes.
$Re[F (\kappa)]$  is close to zero for most of the
frequencies, and only at certain frequencies we see sharp
resonances. In this regime the system acts a {\it window}
for these frequencies.

Another noteworthy observation is that, on one hand, as long as
$\alpha > \alpha_c = 11.64$ (Figs. 2 and 3),
$Re[F (\kappa)]$ is always
greater than its initial value, i.e. $Re[F (\kappa)] > 1$, and
for large frequencies it reaches a ceratin asymptotic value.
On the other hand, when  $\alpha < \alpha_c$ (Figs. 4-6),
we observe an
overall decrease in $Re[F (\kappa)]$ with the increase in
frequency, i.e.,  $Re[F (\kappa)] < 1$ for all $\kappa$'s.

In Fig. 7 we have also studied the behavior of
the dimensionless dynamic permeability
$K^*(\omega_*)$ as  a function of $\omega_*$, the dimensionless
frequency defined as $\omega_*=t_m \omega$, for the case
when $\alpha =0.1$.
Since $\alpha < \alpha_c$, we observe
sharp resonances in the dynamic permability at certain frequences,
which can be explained by the non-Newtonian behavior of the
fluid.

\section{discussion}

In this paper we have studied
non-Newtonian effects on the dynamics of fluids in porous
media by calculating a flow of  Maxwell fluid in a tube,
which oscillates longitudinally and is subject to an oscillatory
pressure gradient. The
present study investigates novelties brought about into
the classic Biot's theory [1] of
propagation of elastic waves in a fluid-saturated
porous solid by inclusion of non-Newtonian effects.
We have used time Fourier transform and converted the
problem to the frequency domain. We have calculated the dynamic
permeability, thus modifying the
work of del Rio, Lopez de Haro, Whitaker [4]
by inclusion of the effect of longitunally oscillating tube walls. We
investigated how
the function $F(\kappa)$, which measures the deviation from
Poiseuille flow friction as a function of frequency parameter $\kappa$,
is modified by non-Newtonian effects.
Our work thus provides  a combined theory of
non-Newtonian flow in a longitudinally
oscillating tube, which constitutes the basis for a realistic
model of the effects of elastic waves in a fluid-saturated
porous space. The
present analisys clearly demonstrates
the existence of a
transition from a dissipative to
elastic regime (as $\alpha$ decreases),
in which sharp enhancements of flow (resonances) occur.

The importance of the current work is two-fold:

(A) We studied modifications brought about
by non-Newtonian effects into Biot's theory.
The investigation of the function $F(\kappa)$ is
important for a number of
applications, since $F(\kappa)$ uniquely determines the
response of a realistic fluid-saturated porous medium to the
elastic waves. Thus, determination of  $F(\kappa)$
for non-Newtonian (Maxwell) fluid  is necessary to guide
e.g., oil-field exploration applications.

(B) As we have seen from Fig. 7, non-Newtonian effects
cause substantial enhancements in dynamic permeability.
We were not able to determine in the literature what
value of $\alpha$ a natural crude oil would have. However,
as it can be seen from Fig.7 for $\alpha=0.1$,
we can get up to 60 times increase in permeability at
certain resonant frequences.
Lower $\alpha$'s yield even more drastic emhancements.
At any rate, we obtained an analytical expression for
$F(\kappa)$ (Eq.(9)), which can provide the behavior of
this function for any given $\alpha$.

In the introduction section, the practical impact of the
possibility of acoustic stimulation of oil reservoirs
has been outlined.
This result clearly demonstrates that in the crude oil,
that can be modelled as a Maxwell fluid,  there are certain
resonant frequencies at which oil production can be
increased significantly if the well is irradiated with
elastic waves at these frequencies.

\acknowledgments

This work was supported by the Iowa State University Center for
Advanced Technology Development and ETREMA Products, Inc.

\centerline{\bf Figure captions}

Fig. 1 Behavior of $Re[F (\kappa)]$ (solid line)
and $Im[F (\kappa)]$ (dashed line),
as function of $\kappa$ according to Eq.(9).
Here, $\alpha = \infty$.

Fig. 2 Same as in Fig. 1 but for $\alpha= 10^4$.

Fig. 3 Same as in Fig. 1 but for $\alpha= 100.0$.

Fig. 4 Same as in Fig. 1 but for $\alpha= 10.0$.

Fig. 5 Same as in Fig. 1 but for $\alpha= 1.0$.

Fig. 6 Same as in Fig. 1 but for $\alpha= 10^{-3}$.

Fig. 7 Behavior of $Re[K^*(\omega_*)] = Re[K(\omega_*) / K(0)]$
(solid line) and
$Im[K^*(\omega_*)] = Im[K(\omega_*) / K(0)]$
(dashed line),
as a function of $\omega_*$ according to Eq.(6).
Here $\alpha=0.1$.

\begin{thebibliography}{}
\bibitem{1}
M.A. Biot, J. Acoust. Soc. Am. , {\bf 28}, 179 (1956);
M.A. Biot, J. Acoust. Soc. Am. , {\bf 28}, 168 (1956)
\bibitem{2}
I.A. Beresnev and P.A. Johnson, Geophys., {\bf 59}, 1000 (1994);
T. Drake and I. Beresnev, The American Oil \& Gas Reporter, September
1999, p.101
\bibitem{3}
C. Chang, Q.D. Nguyen, H.P. Ronningsen, J. Non-Newtonian Fluid Mech.,
{\bf 87}, 127 (1999);
B.P. Williamson, K. Walters, T.W. Bates, R.C. Coy and A.L. Milton,
J. Non-Newtonian Fluid Mech., {\bf 73}, 115 (1997);
G.A. Nunez, G.S. Ribeiro, M.S. Arney, J. Feng and D.D. Joseph,
J. Rheol., {\bf 38(5)}, 1251 (1994);
L.T. Wardhaugh and D.V. Boger, J. Rheol., {\bf 35(6)}, 1121 (1991)
\bibitem{4}
J.A. del Rio, M. Lopez de Haro and S. Whitaker, Phys. Rev., {\bf E58},
6323 (1998)
\bibitem{5}
D.L. Johnson, J. Koplik and R. Dashen, J. Fluid Mech., {\bf 176},
379 (1987);
M.-Y. Zhou and P. Sheng, Phys. Rev. {\bf B39}, 12027 (1989);
M. Avellaneda and S.T. Torquato, Phys. Fluids A {\bf 3}, 2529 (1991);
M. Sahimi, Rev. Mod. Phys, {\bf 65}, 1393 (1993);
P.Sheng and M.-Y. Zhou, Phys. Rev. Lett., {\bf 61}, 1591 (1998)
\end{thebibliography}
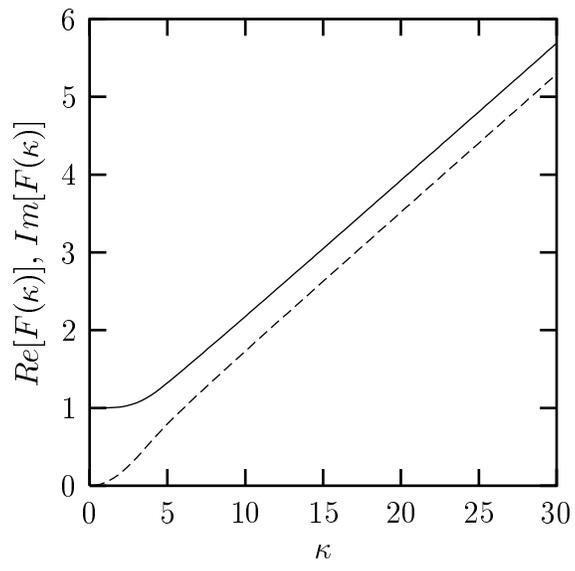
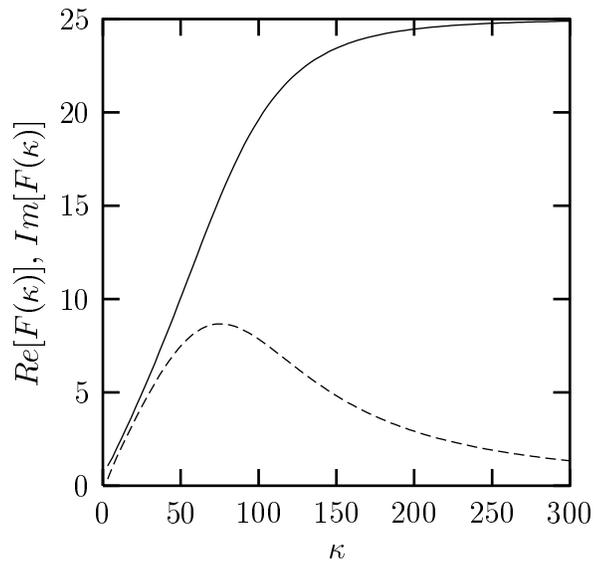
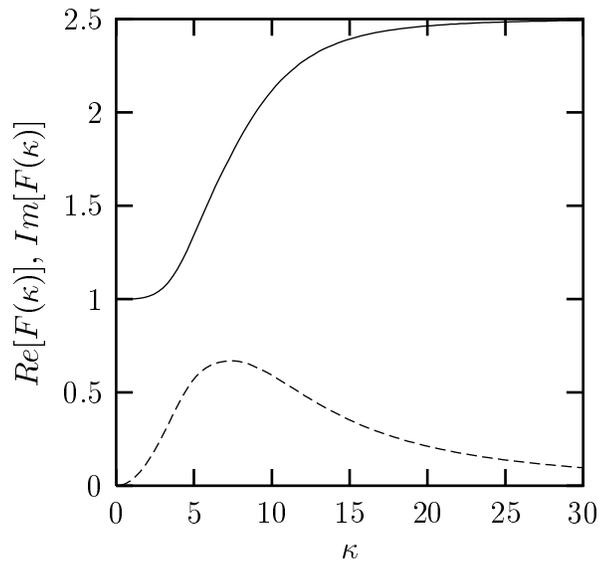
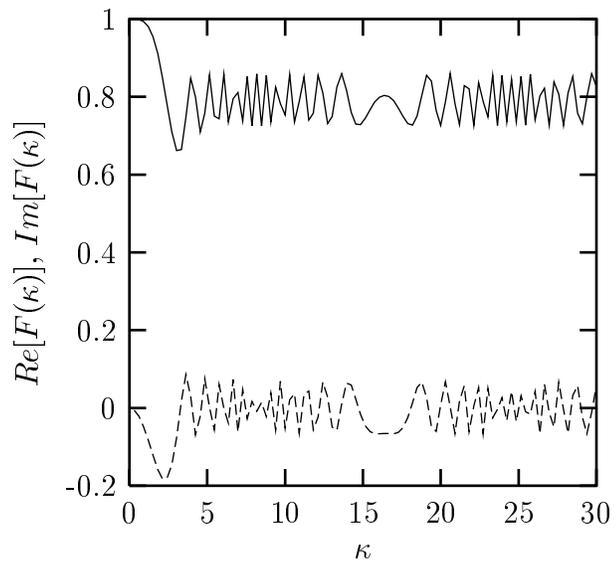
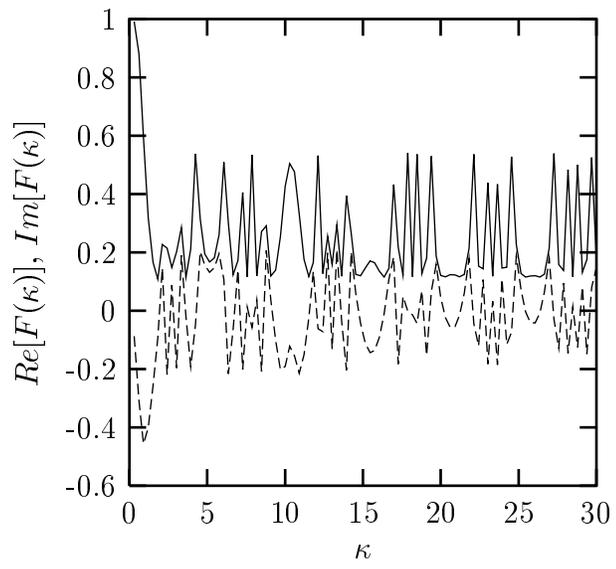
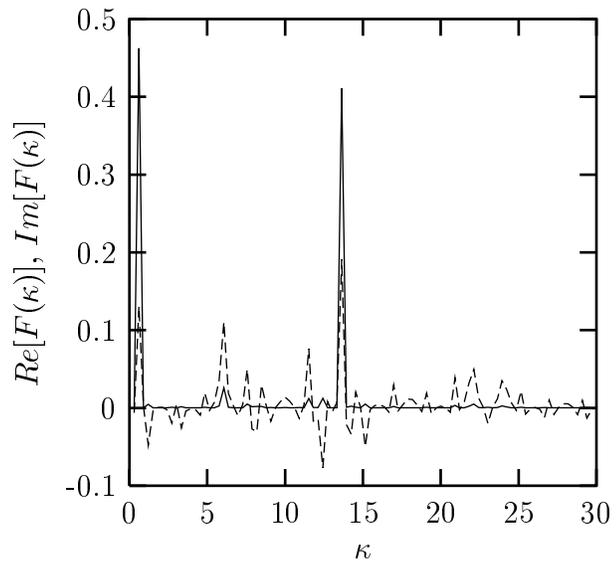
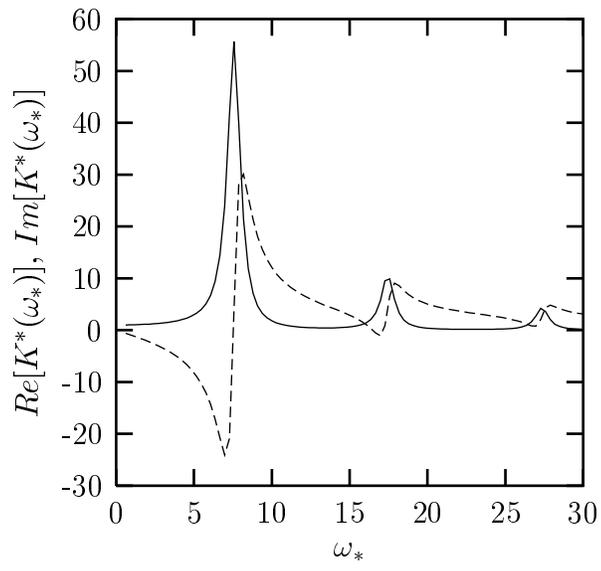
\end{document}